# Neutron Scattering Study of Fluctuating and Static Spin Correlations in the Anisotropic Spin Glass Fe$_2$TiO$_5$

Yu Li[1], P. G. LaBarre[2], D. M. Pajerowski[3], A. P. Ramirez[2], S. Rosenkranz[1], and D. Phelan[1]

[1]*Materials Science Division, Argonne National Laboratory, Lemont, IL 60439, USA*

[2]*Physics Department, University of California Santa Cruz, Santa Cruz, California 95064, USA*

[3]*Neutron Scattering Division, Oak Ridge National Laboratory, Oak Ridge, TN 37830, USA*

**Abstract**

The anisotropic spin glass transition, in which spin freezing is observed only along the ***c***-axis in pseudobrookite Fe$_2$TiO$_5$, has long been perplexing because the Fe$^{3+}$ moments ($d^5$) are expected to be isotropic. Recently, neutron diffraction demonstrated that surfboard-shaped antiferromagnetic nanoregions coalesce above the glass transition temperature, $T_g \approx 55$ K, and a model was proposed in which the freezing of the fluctuations of the surfboards' magnetization leads to the anisotropic spin glass state. Given this new model, we have carried out high resolution inelastic neutron scattering measurements of the spin-spin correlations to understand the temperature dependence of the intra-surfboard spin dynamics on neutron (picosecond) time-scales. Here, we report on the temperature-dependence of the spin fluctuations measured from single crystal Fe$_2$TiO$_5$. Strong quasi-elastic magnetic scattering, arising from intra-surfboard correlations, is observed well above $T_g$. The spin fluctuations possess a steep energy-wave vector relation and are indicative of strong exchange interactions, consistent with the large Curie-Weiss temperature. As the temperature approaches $T_g$ from above, a shift in spectral weight from inelastic to elastic scattering is observed. At various temperatures between 4 K and 300 K, a characteristic relaxation rate of the fluctuations is determined. Despite the freezing of the majority of the spin correlations, an inelastic contribution remains even at base temperature, signifying the presence of fluctuating intra-surfboard spin correlations to at least $T/T_g \approx 0.1$ consistent with a description of Fe2TiO5 as a hybrid between conventional and geometrically frustrated spin glasses.

## I. Introduction

Spin glasses[1] are systems in which magnetic moments freeze below $T_g$ into configurations, which can possess short-range order but that also possess inherent randomness and lack long-range order. The formation of spin glasses is usually associated with a characteristic rugged landscape of metastable states resulting from the interplay of random disorder and frustration, the precise precise roles of which are still of significant debate[2,3,4,5,6]. Experimentally, the observed $T_g$ of spin glasses depends upon the time-scale of the experimental probe because the energies of the spin dynamics are strongly temperature-dependent, particularly in the vicinity of $T_g$. Frequently the frequency dependence of $T_g$, measured via AC magnetic susceptibility (1 to $10^5$ Hz) is modeled either by a Vogel-Fulcher relation [7] or a dynamical scaling law [8]. Inelastic neutron scattering experiments, on the other hand, measure the Fourier reciprocate of a spin-spin correlation function on the much faster time scale of picoseconds ($\sim 10^{12}$ Hz). Additionally, unlike AC susceptibility, neutron scattering probes spin-spin correlations which give rise to peaks at specific wave-vectors. Previous reports of inelastic neutron scattering on conventional spin glasses such as Cu-Mn[9,10], Ni-Mn[11] and Y(Mn$_{0.9}$Al$_{0.1}$)$_2$[12] indicated that the temperature-dependent spin relaxation rates, extracted from the energy-widths of the observed cross-section, can similarly be described by a Vogel-Fulcher law. Nevertheless, whereas Vogel-Fulcher predicts spin-spin correlation lifetimes to diverge at $T_g$, the observed behavior remains poorly understood. For instance, in some conventional systems such as Cu-Mn, this divergence does not occur, but rather temperature-independent relaxation rates were observed below $T_g$ [9]. On the other hand, in some spin glasses which exhibit strong frustration and are thus considered to be unconventional, such as in Y$_2$Mo$_2$O$_7$[13], CeNi$_{0.4}$Cu$_{0.6}$[14], (Ni$_{0.4}$Mn$_{0.6}$)TiO$_3$[15], and BaFe$_{2-x}$Ni$_x$As$_2$[16], only 'elastic' signals were observed below $T_g$, consistent with completely frozen spins at the base temperature within the instrumental resolution. The study of spin glasses that exhibit unique or anomalous behavior may therefore lead to a more advanced understanding of the dynamics and spin-freezing processes.

In light of the different behaviors of conventional and strongly frustrated spin glasses, a particularly anomalous system is the pseudobrookite compound, $Fe_2TiO_5$ which possesses an anisotropic spin glass transition – below $T_g \approx 55$ K, a cusp in susceptibility is seen only along the *c*-axis, with no visible anomaly along *a* or *b*[17,18,19,20,21,22,23]. This is puzzling because $Fe^{3+}$ is an isotropic *s*-state ($d^5$) without single-ion anisotropy. Furthermore, Furthermore, the Fe3+ ions are randomly located on the A and B sites of the pseudobrookite structure, similar to A-B mixing in inverse spinels, which further reduces the likelihood of, say, an interaction-induced anisotropy. Since the Weiss temperature, $\theta_{CW} = \sim -900$ K, indicates the presence of strong antiferromagnetic interactions with significant frustration, $\theta_{CW}/T_g \approx 16$, $Fe_2TiO_5$ is clearly in the class of strongly frustrated magnets. Our recent neutron diffraction study of the spin correlations in this system revealed the presence of strong diffuse scattering indicative of nanoscale order [24]. We found that strong geometrical frustration limits correlations along the *b*-axis to nearest neighbors only. Furthermore, evidence of nanoscale surfboard-shaped regions was observed, with the surfboards developing at temperatures well above $T_g$. Within the surfboards, the magnetic moments are aligned parallel or antiparallel to the *a*-axis (i.e., perpendicular to the direction where spin freezing is observed in susceptibility). The magnetic $Fe^{3+}$ and nonmagnetic $Ti^{4+}$ cations are understood to be "randomly" distributed on the two different crystallographic sites in $Fe_2TiO_5$. The nonmagnetic $Ti^{4+}$ can thus be thought of as spin vacancies. The anisotropic spin freezing was then understood as a consequence of a fluctuation-induced inter-surfboard interaction, i.e., a purely magnetic version of the van der Waals force[24].

While the formation of surfboards begins on cooling at $T > 5T_g$, it should be noted that this is not an example of a Griffith's phase since the AF intra-surfboard order parameter is qualitatively different from the ensuing spin glass forming at $T_g$. Our previous neutron diffraction measurements [24], performed on Corelli [25], provided both "elastic" and "total" intensities, where the "elastic" intensity is measured within an ≈1 meV energy resolution and the total represents the summation over both elastic and inelastic scattering contributions. Such analysis hinted at the presence of varying static and dynamic spin correlations over a wide temperature range; however, to properly probe their static vs dynamic nature, true inelastic neutron scattering measurements are required. Only the results of limited inelastic measurements on $Fe_2TiO_5$, which were interpreted within the context of Vogel-Fulcher scaling, have been reported in the literature [22,23]. Given the much more recent observation of the surfboards, more comprehensive measurements are needed.

Here we report the results of inelastic neutron scattering measurements of a $Fe_2TiO_5$ single crystal at a number of temperatures above and below $T_g$. Spin fluctuations resulting in quasi-elastic magnetic neutron scattering are observed well above $T_g$. The quasi-elastic scattering shows a strongly anisotropic wave-vector transfer ($Q$) dependence and is peaked at the positions in $Q$-space where we previously reported evidence of surfboards in diffraction experiments. No change in the position of the wave-vector of magnetic scattering is observed with increasing energy, indicating a very steep dispersion relation due to strong Fe-Fe exchange interactions. As the temperature is decreased towards $T_g$, a continuous slowing down of the fluctuations and a transfer of spectral weight from quasi-elastic to elastic scattering is observed, indicating spin freezing on the THz timescales. Nevertheless, quasi-elastic scattering remains observed down to $T/T_g \approx 0.1$, signifying a remnant presence of spin fluctuations[26] in contrast to a completely frozen spin scenario. In addition, fluctuations at heretofore un-observed wave-vectors are also observed and are consistent with a slight canting of moments. Thus $Fe_2TiO_5$ exhibits features of both conventional as well as geometrically frustrated spin glass.

## II. Details of Experiments and Data Analysis

The single crystal of $Fe_2TiO_5$ was grown by J. P. Remeika (deceased), formerly of Bell Labs and comes from the same collection which we have previously characterized [24]. Inelastic neutron scattering measurements were performed on the Cold Neutron Chopper Spectrometer (CNCS)[27] at the Spallation Neutron Source, Oak Ridge National Laboratory (Oak Ridge, TN USA). Empty can measurements were subtracted as a background. We used standard data reduction routines to transform data to physical coordinates with Mantid[28]. Experimentally, we define $Q=k_i-k_f$ and $\hbar\omega =E_i-E_f$, where $k_i$ and $k_f$ are the incident and scattered wave-vectors and $E_i$ and $E_f$ are the incident and scattered neutron energies. The sample has an orthorhombic lattice with $a$, $b$, $c$ approximately 3.73, 9.31, and 10.07 Å respectively, and was aligned with $c$ normal to the horizontal scattering plane thus providing only limited detector coverage along $L$. Unless otherwise specified, the values of $L$ throughout this work are 0 by default. Measurements were performed with $E_i =3.32$ meV or $E_i =12$ meV, and 4-dimensional ($Q$ and $\hbar\omega$) volumes of intensity were constructed from a series of rotations of the sample around the axis normal to the horizontal scattering plane. The energy resolution at the elastic line was 0.1 and 0.7 meV full-width-half-maximum for $E_i =3.32$ and 12 meV, respectively[29].

In order to separate the elastic and quasi-elastic (inelastic) contributions to the $\hbar\omega$-dependence of neutron scattering spectra appropriately, we consider that the measured intensity represents a convolution of the intrinsic scattering function, $S(Q,\hbar\omega)$ with a Gaussian resolution function, $g(\hbar\omega,\Gamma_{res})$, of width $\Gamma_{res}$.

$$I(\mathbf{Q}, \hbar\omega) = \int_{-\Delta}^{\Delta} S(\mathbf{Q}, \hbar\omega) \times g(\hbar\omega - \hbar\omega', \Gamma_{res}) d(\hbar\omega')$$

Here, $\Delta$ is a cut-off energy much larger than $\Gamma_{res}$.

**III. Results and Discussion**

In Fig. 1, we present the measured elastic ($\hbar\omega = 0$) neutron scattering patterns in the [$H,K,0$] plane at $T$ = 1.5 K, 55 K, 100 K, 200 K, and 300 K with $L$ = [-0.3,0.3] reciprocal lattice unit (r.l.u.) for $E_i$ = 12 meV and $T$ = 1.5 K, 100 K, 200 K, and 300 K with $L$ = [-0.2,0.2] r.l.u. for $E_i$ = 3.32 meV. The intensities are integrated over an energy range smaller than the instrumental resolution. Streaks of scattering that are narrow in $H$ and extended in $K$ are observed at $H=n+½$, $n=0$, $\pm 1, \pm 2$ ... at and below 100 K. This reflects the formation of surfboards as we described previously[24]. As we will discuss below, the scattering within this "elastic" window consists of a true elastic contribution (within the resolvable energy of the instrument) as well as the central part of a broadened quasi-elastic component, which is inherently inelastic. Besides these sharp streaks, there are fainter and broader blobs of intensities centered at the same location in reciprocal space. These blobs lack a significant temperature-dependence and indeed remain at higher temperatures ($T$>100 K) [Fig1(c-e,h,i)], suggesting that they may be structural rather than magnetic in nature and likely arise from underlying structural disorder due to the site mixing of $Fe^{3+}$ and $Ti^{4+}$ cations.

Figure 2 shows inelastic neutron scattering intensity, S($\mathbf{Q},\hbar\omega$), in the [$H, K$] plane at $L$=0 at two characteristic $\hbar\omega$, 2 and 6 meV, for a series of temperatures. At both values of $\hbar\omega$, the shape of the intensity mimics the low-temperature elastic scattering pattern. This immediately indicates that the correlations being probed are intra-surfboard as the nature of the correlation is implicated by its $\mathbf{Q}$-dependence. This inelastic scattering persists up to the highest measured temperature (300

K), demonstrating the existence of fluctuating intra-surfboard correlations far above $T_g$. These fluctuations are gradually enhanced upon cooling towards $T_g$. Below $T_g$, the intensities are dramatically suppressed due to spin freezing and transfer of the spectral weight into the elastic channel; however, there are remnant intensities as shown in Fig.2(a) and (f), indicating that there are remaining fluctuating intra-surfboard correlations at the lowest temperature we can reach. Furthermore, as seen in a color image of these residual spin fluctuations in Fig.3(a), the steepness of these fluctuations as a function of $H$ and $\hbar\omega$ is clearly seen. In Fig.3(b), the momentum dependence of these spin fluctuations along the $H$ at a series of energies is presented. A Lorentzian fit does not exhibit any broadening of momentum linewidth as energy increases, indicating the existence of large local exchange coupling. We note here that no gap is observed at 1.5 K, consistent with the absence of either single-ion anisotropy or long range order.

In order to gain further insight into the temperature-dependence of the intra-surfboard spin dynamics, we present in Fig. 4, energy spectra at constant $\boldsymbol{Q} = (0.5, 1.5)$ at a series of temperatures where the spectra are dominated by the intensity arising from intra-surfboard correlations. We fit these energy-dependent spectra to the sum of a purely elastic component, given by a delta function scaled by an amplitude, $f_0 \delta(\hbar\omega)$, and a quasielastic component. The amplitude $f_0$ is proportional to the strength of the purely elastic component. For the quasi-elastic component, we consider that the scattering function corresponds to that of a damped zero-energy mode. The resulting $S(\hbar\omega)$ is given by:

$$S(\hbar\omega) = f_0 \delta(\hbar\omega) + \frac{\chi_0 \Gamma \hbar\omega}{(\hbar\omega)^2 + \Gamma^2} \times \frac{1}{1 - exp\left(\frac{-\hbar\omega}{k_B T}\right)}$$

Here, $\Gamma$ is a measure of the relaxation rate and inversely proportional to the lifetime of the zero-energy mode, and $\chi_0$ is a measure of the staggered susceptibility. The amplitude of the elastic contribution, $f_0$, as function of temperature, is shown in Fig. 3(b). As the temperature decreases all the way down to the base temperature, a substantial increase of the elastic contribution is clearly observed at $T_g$. The rapid increase in this contribution mirrors the increase of correlation length and indicates that the freezing of intra-surfboard correlations is strongly connected to the spin glass transition, even though the latter is dominated by inter-surfboard correlations. The slight increase of the intensity, $\chi_0$, at 100 K and above may come from a background of elastic nuclear contribution. In Fig. 4(c), the temperature dependence of $\Gamma$ which is obtained from the quasielastic contribution is shown. A reduction of $\Gamma$ on cooling, estimated from the measurement with $E_i$ = 12meV, starts to happen well above $T_g$, from 4.6 meV at 300 K to 2.3 meV at 55 K, similar to other spin glasses and is interpreted as a characteristic of spin freezing. However, due to the limited points above $T_g$, we are not able to determine whether the Vogel-Fulcher scaling rule is violated or not. We note that the value of $\Gamma$ at low temperatures is still finite and significantly larger than the instrumental resolution, suggesting existence of dynamics below $T_g$. The staggered susceptibility $\chi_0(\mathbf{Q})$ with $\mathbf{Q}$= (0.5,1.5,0) displayed in Fig. 4(d), shows a maximum at $T_g$, indicating enhanced AFM correlations on approaching $T_g$.

In Fig. 5 (a-d) we show the Q-dependent scattering intensites, integrated over a small energy range smaller than the data of Fig. 2, and deep within the quasielastic peak. As already hinted at in Fig. 2, we clearly see diffuse maxima, which were unobserved in our previous neutron diffraction measurements, which revealed only a broad background at the same positions. The reason for this discrepancy is that these maxima are resolved at only non-zero $\hbar\omega$. This is exemplified in Fig. 5(a-d), where a series of 2-dimensional images of inelastic neutron scattering intensities in the [$H,K$,0] plane measured are displayed for $\hbar\omega$ = 0.5 meV with $E_i$ = 3.32 meV at different temperatures. In addition to the strong antiferromagnetic spin fluctuations at half-integer

*H*, discussed above, the additional excitations with prolate profiles along the *K* direction at integer *H*, such as at $Q$ = (1,0,0), are observed. As seen in Fig. 5(f), these excitations are only observable below $\hbar\omega$ = 2 meV, and exhibit the highest intensities at the highest measured temperature (300 K) [Fig.5(d)] in contrast to the fluctuations at half-integer *H* which are maximal close to $T_g$, and are strongly suppressed at base temperature (Fig. 5(e)). It is known that $Q$ = (1,0,0) is a forbidden wave vectors for structural Bragg peaks in the ***Cmcm*** space group. Therefore, these intensities are not due to acoustic phonons. Moreover, the observation of these fluctuations at low |$Q$| also suggests that they are magnetic. Note that if spins point along ***a***, as found in our previous work [24], magnetic intensity should be forbidden at $Q$=(1,0,0) because neutrons are only sensitive to the component of spin perpendicular to $Q$. Thus the present observation suggests that there is a weak spin component perpendicular to the ***a***-axis which are not accounted for in the previous model. We have performed a calculation of the magnetic structure factor based on the surfboard model in a previously proposed in a supercell but with the addition of a ***c***-axis canting within the surfboard . We found that such canting could reasonably account for these new observations.  Let us note that while the ***a***-axis moments are still completely frustrated between neighboring double spin chains as proposed previously, the presence of ***c***-axis canting moments could slightly relax the frustration to certain extent that the correlation length along the ***b***-axis is not perceivably affected. Therefore, we suggest that the canting away from the ***a***-axis is the origin of this feature at integer ***H***. However, due to its weak signal and the presence of strong background, whether there exists elastic contribution or spin freezing at this wave vector is not conclusive. Nevertheless, the differing temperature-dependence with respect to the other fluctuating moments is not yet understood.

**IV. Summary**

It is useful to summarize the present results by forming a pseudo-phase diagram that illustrates the temperature-dependence of the spin correlations in $Fe_2TiO_5$ as shown in Fig. 6. At elevated temperatures (Region III), the magnetic susceptibility is Curie-Weiss like and isotropic, reflecting paramagnetic behavior of spins which have instantaneous nearest-neighbor antiferromagnetic correlations, but with no extended, surfboard correlations. We surmise that the appearance of fluctuating intra-surfboard correlations coincide with the deviation from the Curie-Weiss law below 400 K in Region II. It is worthy of mentioning that the extra spin fluctuations at $Q = (1,0,0)$ also appear in this temperature range. Here the intra-surfboard correlations are fluctuating and they slow as the temperature is lowered and the relaxation rate decreases. The susceptibility shows no appreciable anisotropy in Region II. However, below 150 K, the system enters Region I where the dynamics of intra-surfboard correlations have appreciably slowed down and anisotropy appears in the susceptibility. At $T_g$, the spectral weight of intra-surfboard correlations is rapidly increasing in the elastic channel, and the fluctuating susceptibility has peaked. If we consider the anisotropic peak in susceptibility at $T_g$ to be a consequence of inter-surfboard interactions, the implication here is that the freezing of the inter-surfboard dynamics is strongly correlated to the freezing of the intra-surfboard dynamics. Nevertheless, well below $T_g$, some intra-surfboard fluctuations remain, as is also indicated by the anisotropic susceptibility[24].

**Acknowledgements:**

Work at Argonne National Laboratory (neutron scattering measurements, analysis and interpretation) was supported by the U.S. Department of Energy, Office of Science, Office of Basic Energy Sciences, Materials Sciences and Engineering Division. A portion of this research used resources at the Spallation Neutron Source, a DOE Office of Science User Facility operated but Oak Ridge National Laboratory. Work at UCSC was supported by the U.S. Department of Energy grant DE-SC0017862.

**References**

[1] K. Binder, and A.P. Young, Rev. Mod. Phys. 58, 801 (1986).

[2] I. Klich, S.-H. Lee, and K. Lida, Nat. Commun. 5: 3497 (2014).

[3] J. Yang, A. Samarakoon, S. Dissanayake, H. Ueda, I. Klich, K. Lida, D. Pajerowski, N. P. Butch, Q. Huang, J. R. D. Copley, and S.-H. Lee, Proceedings of the National Academy of Sciences, 112, pp.11519-11523 (2015).

[4] Leon Balents, Nature 464, 199-208 (2010).

[5] Jason S. Gardner, Michel J.P. Gingras, and John E. Greedan, Rev. Mod. Phys. 82, 53 (2010).

[6] A.P. Ramirez, Nature 421, 483 (2003).

[7] S. Shtrikman, and E.P. Wohlfarth, Phys. Lett. A 85, 467-470 (1981).

[8] C. Pappas, F. Mezei, G. Ehlers, P. Manuel, and I.A. Campbell, Phys. Rev. B 68, 054431 (2003).

[9] A.P. Murani, J. Appl. Phys. 49, 1604 (1978).

[10] A.P. Murani, J.L. Tholence, Solid State Commun. 22, 25-28 (1977).

[11] B. Hennion, M. Hennion, F. Hippert, and A.P. Murani, J. Phys. F: Met. Phys. 14, 489 (1984).

[12] K. Motoya, T. Freltoft, P. Boni, and G. Shirane, Phys. Rev. B 38, 4796 (1988).

[13] J.S. Gardner, B.D. Gaulin, S.-H. Lee, C. Broholm, N.P. Raju, and J.E. Greedan, Phys. Rev. Lett. 83, 211 (1999).

[14] J.C. Gomez Sal, J. Garcia Soldevilla, J.A. Blanco, J.I. Espeso, J. Rodriguez-Fernandez, F. Luis, F. Bartolome, and J. Bartolome, Phys. Rev. B 56, 11741 (1997).

[15] R.S. Solanki, S.-H. Hsieh, C.H. Du, G. Deng, C.W. Wang, J.S. Gardner, H. Tonomoto, T. Kimura, and W.F. Pong, Phys. Rev. B 95, 024425 (2017).

[16] Xingye Lu, David W. Tam, Chenglin Zhang, Huiqian Luo, Meng Wang, Rui Zhang, Leland W. Harriger, T. Keller, B. Keimer, L.-P. Regnault, Thomas A. Maier, and Pengcheng Dai, Phys. Rev. B 90, 024509 (2014).

[17] U. Atzmony, E. Gurewitz, M. Melamud, H. Pinto, H. Shaked, G. Gorodetsky, E. Hermon, R.M. Hornreich, S. Shtrikman, and B. Wanklyn, Physical Review Letters 43, 782 (1979).

[18] Y. Yeshurun, I. Felner, and B. Wanklyn, Phys. Rev. Lett. 53, 620 (1984).

[19] Y. Yeshurun, and H. Sompolinsky, Phys. Rev. B 31, 3191 (1985).

[20] J.K. Srivastava, W. Treutmann, and E. Unterstelller, Phys. Rev. B 68, 144404 (2003).

[21] J. Rodrigues, W.S. Rosa, M.M. Ferrer, T.R. Cunha, M.J.M. Zapata, J.R. Sambrano, J.L. Martinez, P.S. Pizani, J.A. Alonso, A.C. Hernandes, and R.V. Goncalves, J. Alloys Compd. 799, 563 (2019).

[22] Y. Yeshurun, J.L. Tholence, J.K. Kjems, and B. Wanklyn, J. Phys. C: Solid State Phys. 18, L483 (1985).

[23] R.L. Lichti, S. Kumar, and C. Boekema, J. Appl. Phys. 63, 4351 (1988).

[24] P.G. LaBarre, D. Phelan, Y. Xin, F. Ye, T. Besara, T. Siegrist, S.V. Syzranov, S. Rosenkranz, and A.P. Ramirez, Phys. Rev. B 103, L220404 (2021).

[25] F. Ye, Y.H. Liu, R. Whitfield, R. Osborn, and S. Rosenkranz, J. Appl. Crystallogr. 51, 315 (2018).

[26] Ping Miao, Rui Wang, Weiming Zhu, Jiajie Liu, Tongchao Liu, Jiangtao Hu, Shuankui Li, Zhijian Tan, Akihiro Koda, Fengfeng Zhu, Erxi Feng, Yixi Su, Takashi Kamiyama, Yinguo Xiao, and Feng Pan, Appl. Phys. Lett. 114, 203901 (2019).

[27] G. Ehlers, A.A. Podlesnyak, and A.I. Kolesnikov, Rev. Sci. Instrum. 87, 093902 (2016).

[28] O. Arnold, J.C. Bilheux, J.M. Borreguero, A. Buts, S.I. Campbell, L. Chapon, M. Doucet, N. Draper, R. Ferraz Leal, M.A. Gigg, V.E. Lynch, A. Markvardsen, D.J. Mikkelson, R.L. Mikkelson, R. Miller, K. Palmen, P. Parker, G. Passos, T.G. Perring, P.F. Peterson, S. Ren, M.A. Reuter, A.T. Savici, J.W. Taylor, R.J. Taylor, R. Tolchenov, W. Zhou, and J. Zikovsky, Nucl. Instrum. 764, 156-166 (2014).

[29] These values were obtained by fitting the elastic line with a Gaussian function and agree with the estimation from PyChop model. https://rez.mcvine.ornl.gov/.

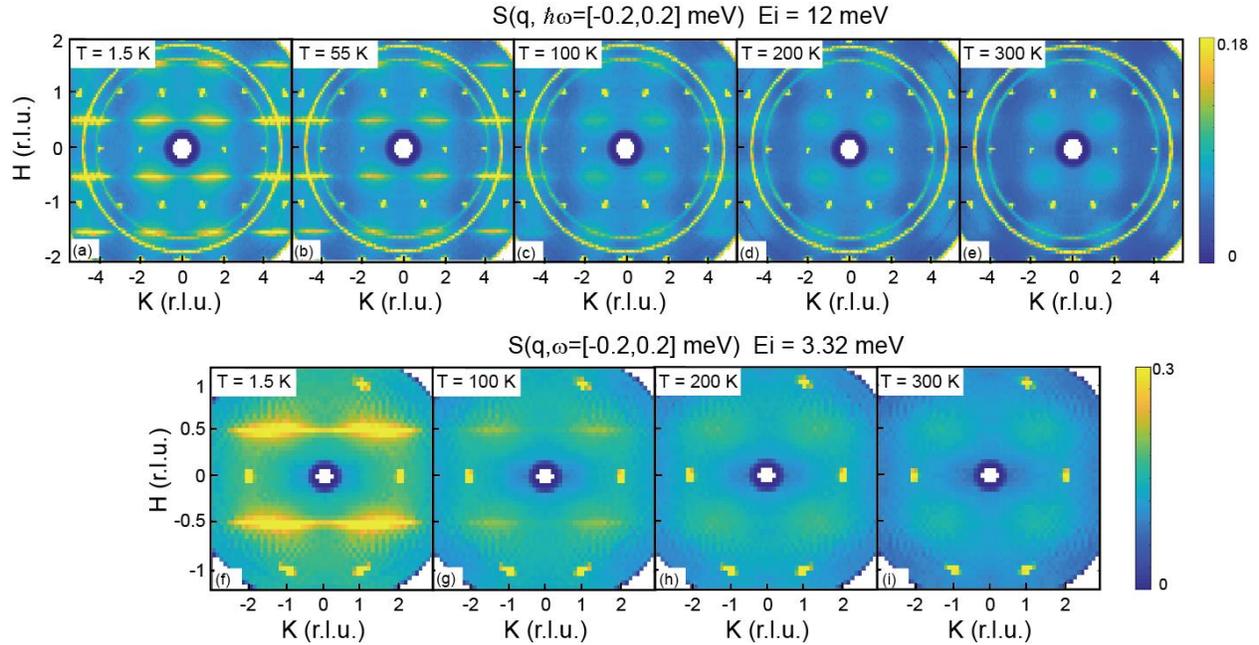

*Figure 1. 2-dimensional images of neutron diffraction intensities in the HK0 plane for different temperatures and incident neutron energies. Bright spots at integer numbers are nuclear Bragg peaks. Narrow streaks below 55 K at $H = n + 0.5, n = 0, \pm1, \pm2$ ...are diffuse scattering associated with the surfboard nanoregion. There are broad temperature-independent intensities at the same location persisting up to 300 K and they reflect the underlying structural disorders. (a-e) Neutron diffraction measured with $E_i$ = 12 meV. (f-i) Neutron diffraction measured with $E_i$ = 3.32 meV.*

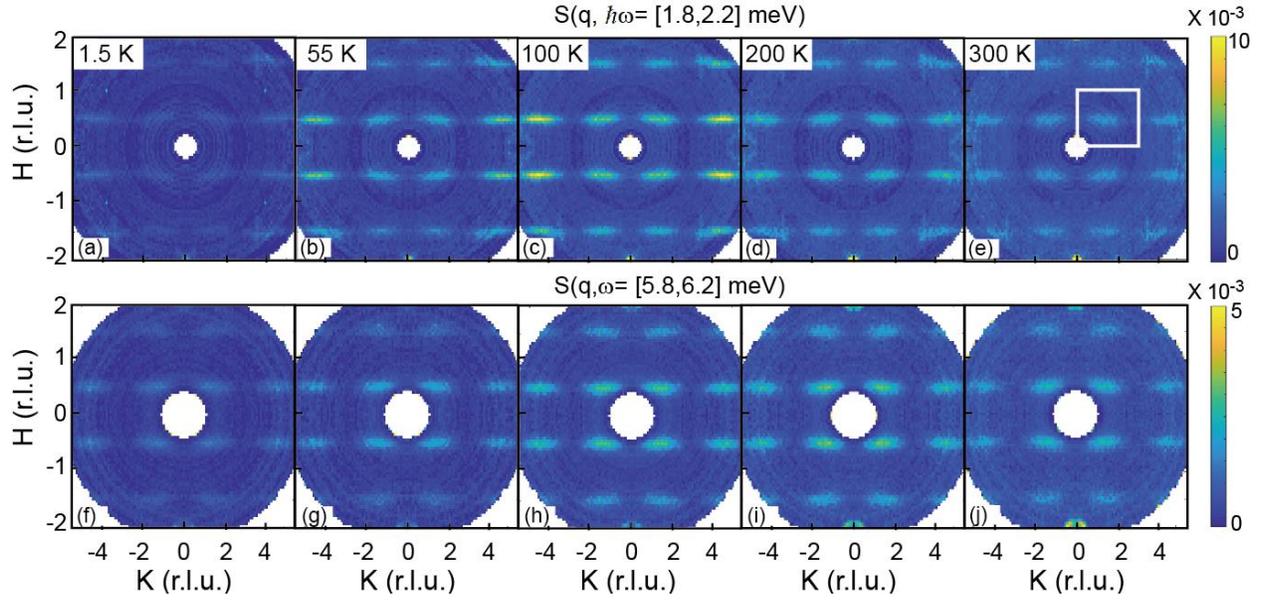

*Figure 2: Inelastic neutron scattering intensities plotted in the [H,K,0] plane at different temperatures and energy transfer. Intensities arising at ($\pm 2,0,0$) are acoustic phonons. (a-e) spin fluctuations measured at $\hbar\omega = 2$ meV for five different temperatures. The white square in € represent an approximate periodicity of these spin fluctuations in the [H,K,0] plane as discussed in the text. (f-j) spin fluctuations measured at $\hbar\omega = 6$ meV.*

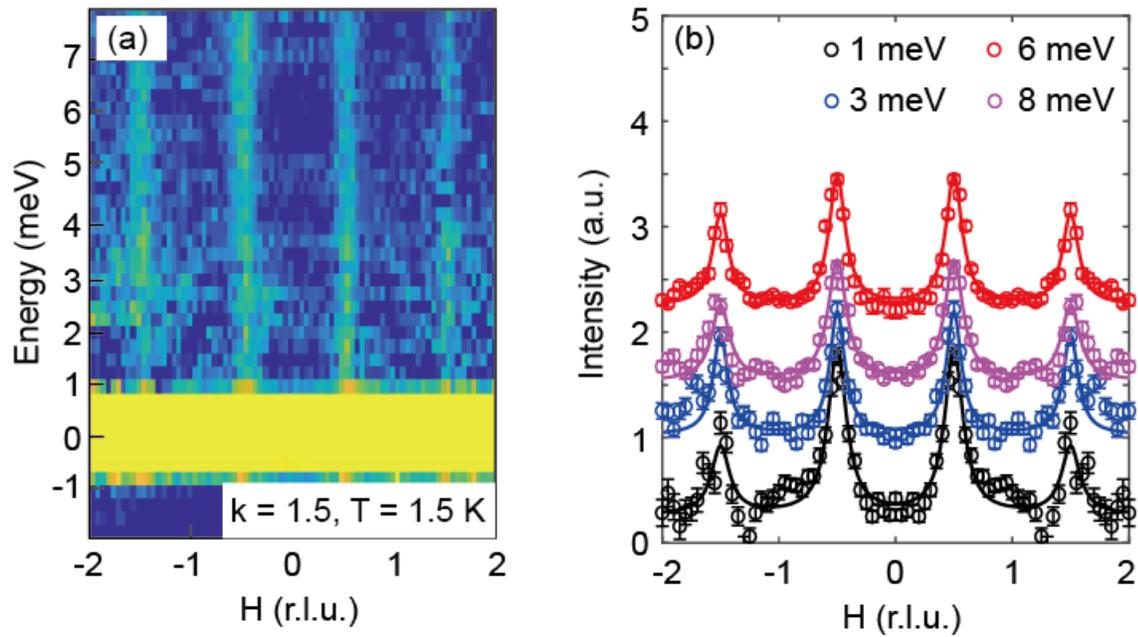

*Figure 3*: *(a) Color image plot of inelastic neutron scattering intensities as a function of H and energy at 2K. Vertical dispersionless fluctuations are clearly observed at half integer H. (b) Cuts along the [H,1.5,0] direction at various energies at 1.5 K. There is a constant shift for each scan along the y-axis. Solid lines are fitting results with a Lorentzian function.*

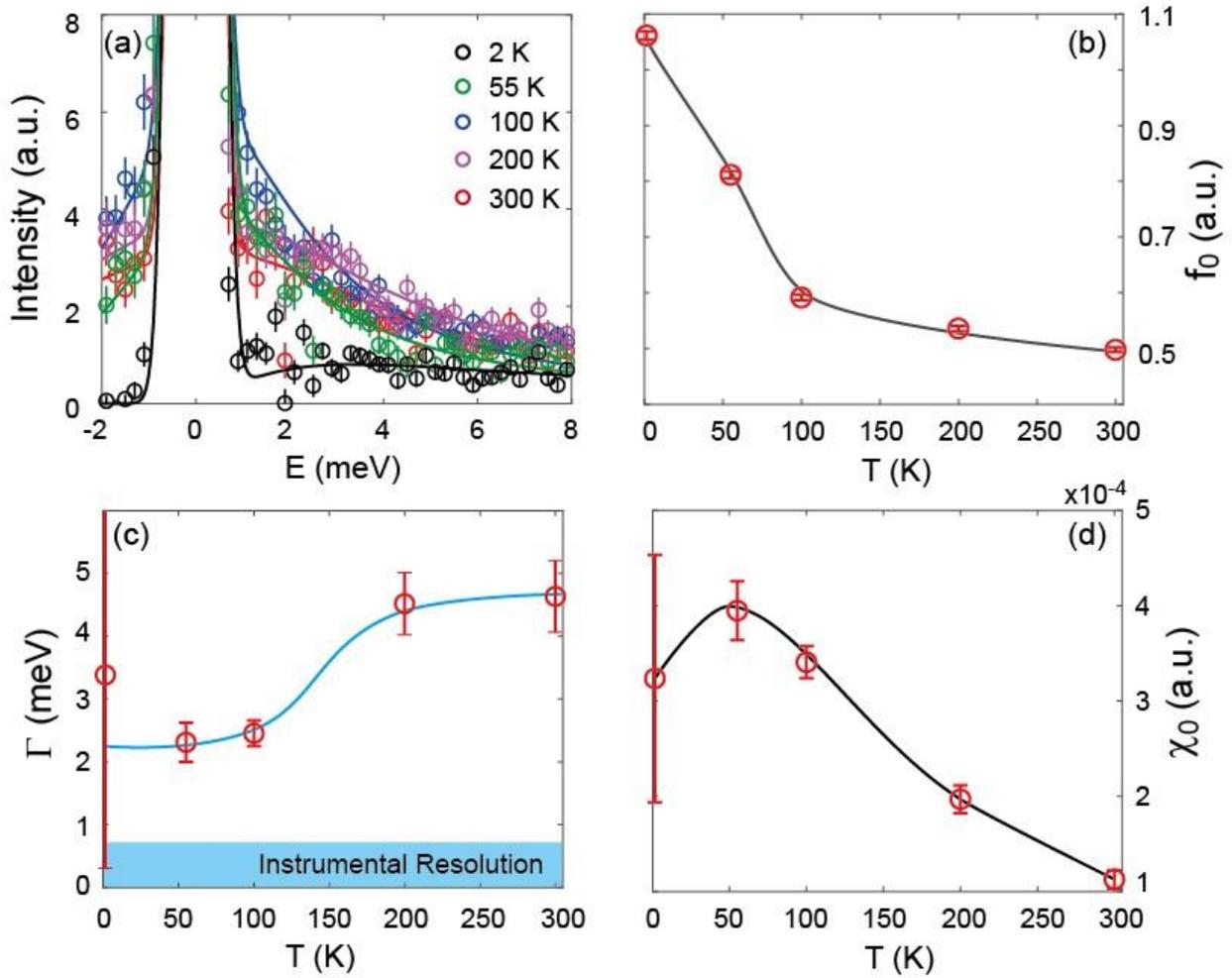

*Figure 4: (a)Constant-**Q** energy spectra of neutron scattering intensities at 5 different temperatures. The solid lines are results from our fitting . (b-d) The estimated $A_{elastic}$, line widths of the Lorentzian function, $\Gamma$ , and the staggered susceptibility,$\chi_0$ (Q) at Q=(0.5,1.5,0), as a function of temperature. The blue shading area represent the instrumental resolution.*

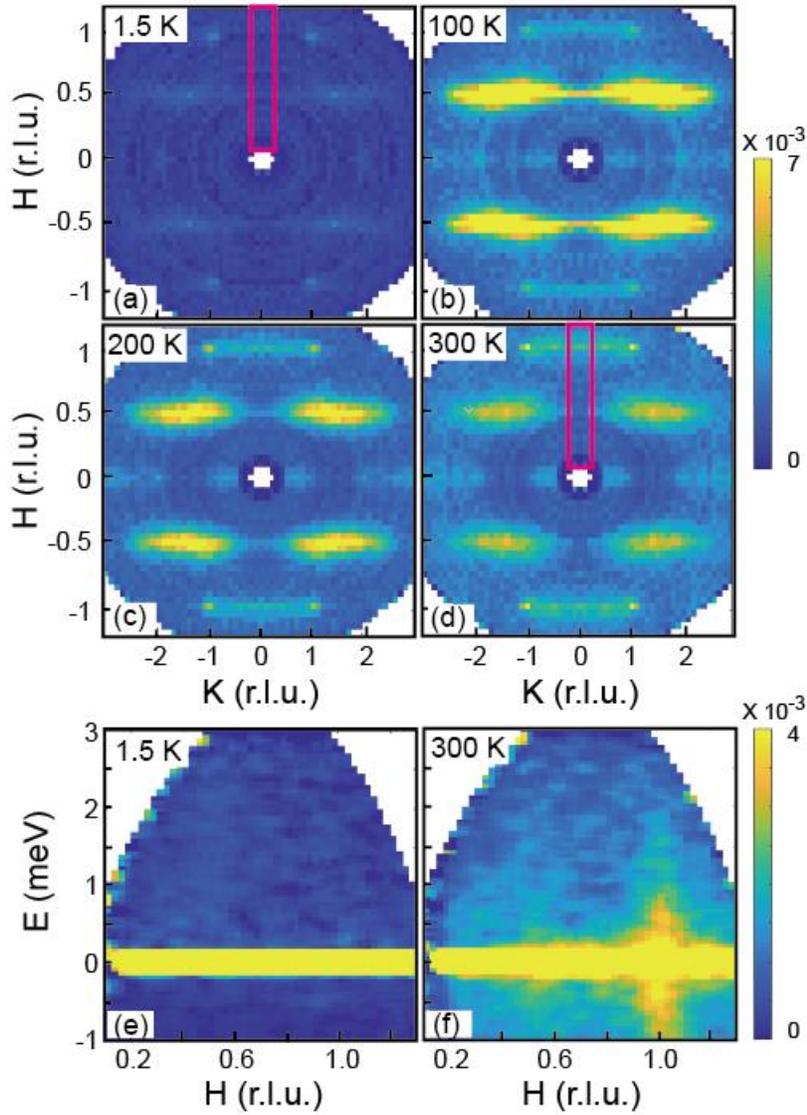

*Figure 5*: *(a-d) Inelastic neutron scattering intensities integrated in the ħω range of [0.3,0.7] meV in the [H,K,0] plane at four different temperatures. (e)(f) Slices of intensities along the direction as marked by the pink rectangles in (a) and (d) as a function of H and energy at the base temperature and 300 K.*

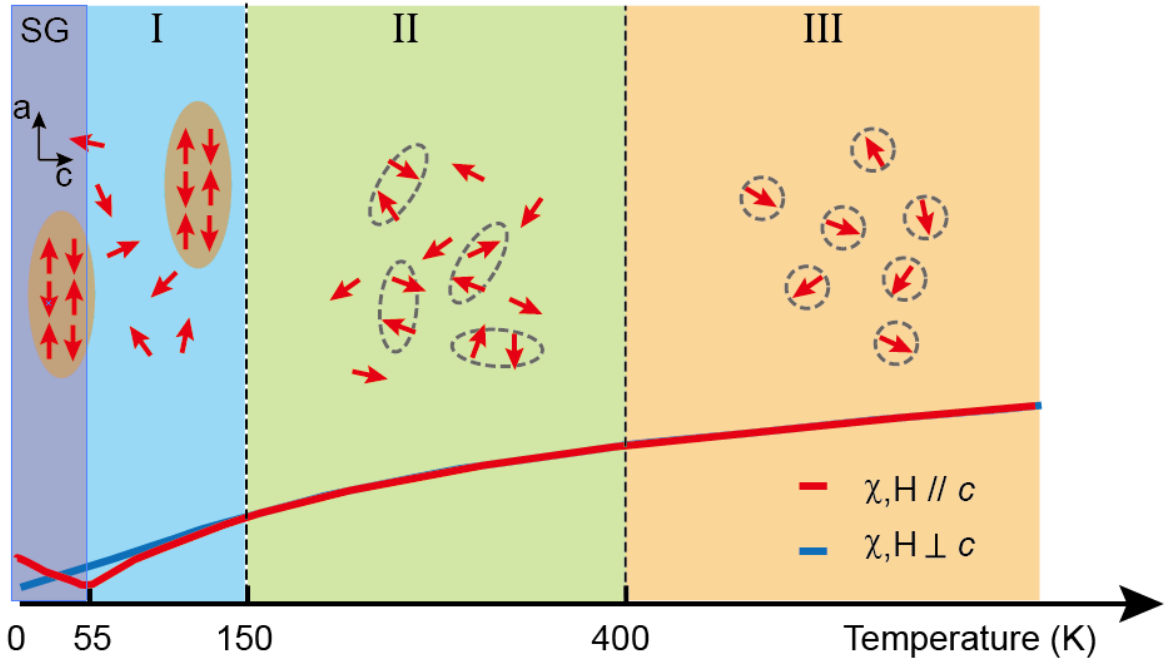

*Figure 6*: *A combined phase diagram determined by magnetic susceptibility and neutron scattering. At high temperatures above 400 K, isolated magnetic moments are not correlated and the susceptibility follow the Curie-Weiss law. Between 150 and 400 K, the deviation from Curie-Weiss law reflect the presence of correlation between magnetic moments. Below 150 K, the surfboard structure is formed and the anisotropy between **a** and **c**-axis start to develop. Below $T_g$~ 55 K, the dynamic surfboard is frozen and become static while the surrounding spin clouds are still dynamically fluctuating.*